\documentclass[aps,prl,groupedaddress,superscriptaddress,longbibliography,
twocolumn,
,amsmath,amssymb]{revtex4-2}
\usepackage{color}
\definecolor{darkblue}{rgb}{0.0, 0.0, 0.55}
\usepackage[colorlinks=true,linkcolor=darkblue,anchorcolor=red,citecolor=darkblue, urlcolor=darkblue]{hyperref}
\usepackage{romannum}
\usepackage{physics}
\usepackage{graphicx}
\usepackage{bm}
\usepackage{ulem}
\usepackage{subfigure}
\usepackage[capitalize]{cleveref}
\usepackage{epstopdf}

\begin{document}

\title{Coexisting of Quantum Hall and Quantum Anomalous Hall phases in Disordered $\mathrm{MnBi_2Te_4}$}
\author{Hailong Li}
\thanks{Hailong Li and Chui-Zhen Chen are co-first authors}
\affiliation{International Center for Quantum Materials, School of Physics,
Peking University, Beijing 100871}
\author{Chui-Zhen Chen}
\thanks{Hailong Li and Chui-Zhen Chen are co-first authors}
\affiliation{School of Physical Science and Technology, Soochow University, Suzhou 215006, China}
\affiliation{Institute for Advanced Study, Soochow University, Suzhou 215006, China}
\author{Hua Jiang}
\email{jianghuaphy@suda.edu.cn}
\affiliation{School of Physical Science and Technology, Soochow University, Suzhou 215006, China}
\affiliation{Institute for Advanced Study, Soochow University, Suzhou 215006, China}
\author{X. C. Xie}
\email{xcxie@pku.edu.cn}
\affiliation{International Center for Quantum Materials, School of Physics,
Peking University, Beijing 100871}
\affiliation{Beijing Academy of Quantum Information Sciences, Beijing 100193, China}
\affiliation{CAS Center for Excellence in Topological Quantum Computation, University of Chinese Academy of Sciences, Beijing 100190, China}
\date{\today }

\begin{abstract}
In most cases, to observe quantized Hall plateaux, an external magnetic field is applied in intrinsic magnetic topological insulators $\mathrm{MnBi_2Te_4}$.
Nevertheless, whether the nonzero Chern number ($C\neq 0$) phase is a quantum anomalous Hall (QAH) state, or a quantum Hall (QH) state, or a mixing state of both is still a puzzle, especially for the recently observed $C=2$ phase [Deng \textit{et al}., Science \textbf{367}, 895 (2020)].
In this Letter, we propose a physical picture based on the Anderson localization to understand the observed Hall plateaux in disordered $\mathrm{MnBi_2Te_4}$.
Rather good consistency between the experimental and numerical results confirms that the bulk states are localized in the absence of a magnetic field and a QAH edge state emerges with $C=1$.
However, under a strong magnetic field, the lowest Landau band formed with the localized bulk states, survives disorder, together with the QAH edge state, leading to a $C=2$ phase.
Eventually, we present a phase diagram of a disordered $\mathrm{MnBi_2Te_4}$ which indicates more coexistence states of QAH and QH to be verified by future experiments.
\end{abstract}


\maketitle

{\textit{Introduction.}}---
Since successfully synthesized in experiments \cite{2Dmater_otrokov2017,cpl_xue2019}, $\mathrm{MnBi_2Te_4}$ as an intrinsic magnetic topological insualtor has attracted great attention \cite{Science_zhang2020,nsr_wangjian2020,natmater_wangyayu2020,prl_wang2019,sciadv_xu2019,nature_otrokov2019,prl_dassarma2020,prx_hao2019}.
Intriguingly, the layered $\mathrm{MnBi_2Te_4}$ is predicted to show plenty of exotic quantum phases, such as Weyl semimetals, quantum anomalous Hall (QAH) insulators, topological axion insulators, etc \cite{prl_wang2019,sciadv_xu2019}.
However, to observe quantized Hall plateaux, it is common to apply a perpendicular magnetic field to raise the interlayer magnetic order of $\mathrm{MnBi_2Te_4}$ \cite{Science_zhang2020,nsr_wangjian2020,natmater_wangyayu2020}.
On one hand, a magnetic field can induce an exchange gap to make it easier to observe the QAH effect \cite{prl_qah2008,science_qah2010,science_qah2013,arp_qah2016}, which gives rise to a higher Chern number with increasing the magnetic field \cite{Science_zhang2020,nsr_wangjian2020}.
On the other hand, the quantum Hall (QH) effect can also exist, in which case a lower Chern number corresponds to a larger magnetic field, opposite to the QAH case.

Usually, the fabricated $\mathrm{MnBi_2Te_4}$ in experiments owns extremely low mobility from 74 $\mathrm{cm}^2\mathrm{V}^{-1}\mathrm{s}^{-1}$ to 1500 $\mathrm{cm}^2\mathrm{V}^{-1}\mathrm{s}^{-1}$, indicating the presence of strong disorder \cite{cpl_xue2019,natmater_wangyayu2020,nsr_wangjian2020,Science_zhang2020}.
It thus suggests that the Anderson localization \cite{physrev_anderson1958} may play a key role in these systems.
In this case, bulk states are localized and a mobility gap may dominate the QAH phase instead of a band gap \cite{prl_nagaosa2007,progtheorphys_hikami1980}.
Although it has been suggested that the localized bulk states can also form Landau bands due to the magnetic-field-induced delocalization \cite{prb_ando1989,prb_azbel1992}, such a process, especially accompanied by the coexistence of QAH and QH states, remains elusive in experiments.
In this regard, the phase transitions of $\mathrm{MnBi_2Te_4}$ under a magnetic field can be full of complexity and also of great interest, especially for the cases with higher Chern numbers.

\begin{figure*}[htbp]
  \includegraphics[width=2\columnwidth]{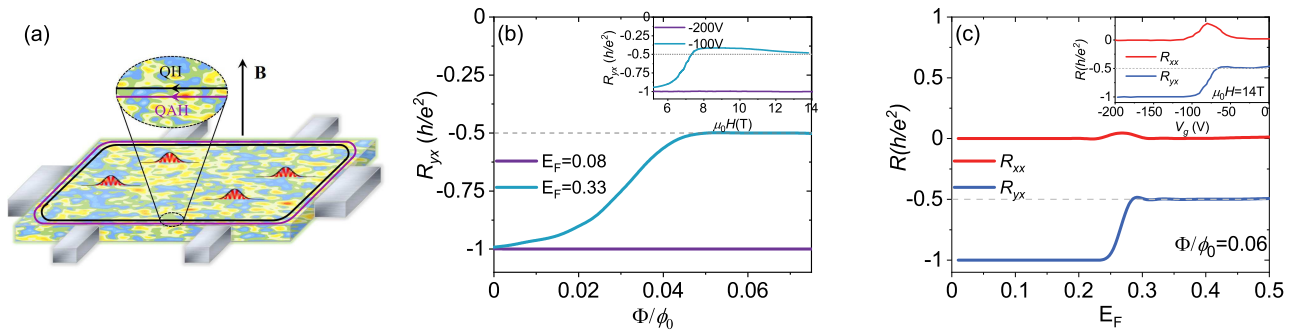}
  \caption{(color online) (a) shows a schematic plot of a six-terminal disordered $\mathrm{MnBi_2Te_4}$ device under a strong magnetic field. The wave packets represent the localized bulk states. In the absence of a magnetic field, bulk states are localized by disorder and the QAH edge state emerges. For a strong magnetic field, these localized bulk states can still form Landau bands coexisting with the QAH state. (b) Calculated Hall resistance $R_{yx}$ as a function of magnetic field $\Phi/\phi_0$ under different Fermi energies.
  (c) Calculated longitudinal resistance $R_{xx}$ and $R_{yx}$ as a function of Fermi energy. The insets show the experimental measured $R_{xx}$ and $R_{yx}$ adopted from Ref. \cite{Science_zhang2020}, where $\mu_0H$ and $V_g$ correspond to the magnetic field and Fermi energy, respectively. The numerical simulation is performed with sample size of $400\times400$ sites and averaged over 320 disorder configurations.}
  \label{Fig1}
\end{figure*}

Recently, the observation of the QAH effect was reported in $\mathrm{MnBi_2Te_4}$ thin flakes \cite{Science_zhang2020}, which implies the existence of Anderson localization in a topological system. The well-developed QAH effect emerges as the Hall resistance $R_{yx}$ reaches $-h/e^2$ under a weak magnetic field. Apart from the QAH plateau, $R_{yx}$ goes through an additional plateau at $-h/(2e^2)$ under a strong magnetic field [see the inset of \cref{Fig1}].
The second Hall plateau is speculated as the emergence of an additional QH edge state in Ref. \cite{Science_zhang2020}.
However, it has not been well understood because the bulk states accounting for the additional Landau band will cover the QAH edge state and lead to an unquantized Hall resistance under the weak magnetic field.
Thus, it is essential to use the Anderson localization to understand the mechanism of both Hall plateaues in a unified framework, thus, help us to identify the complex phase transitions in disordered $\mathrm{MnBi_2Te_4}$.

In this Letter, we provide a physical picture based on the Anderson localization to understand the complex phase transition from a pure QAH phase to a phase composing of QAH and QH edge states in disordered $\mathrm{MnBi_2Te_4}$ [see \cref{Fig1}(a)].
Via an established disordered Hamiltonian and the nonequilibrium Green's function method \cite{zpb_mackinnon1985,prb_jauho1994,arxiv_datta2020}, we calculate the Hall  resistance $R_{yx}$ and the longitudinal resistance $R_{xx}$ of a six-terminal device [see \cref{Fig1}].
The results are in good agreement with the experimental data [see \cref{Fig1}(b) and 1(c)].
Furthermore, we capture the underlying physics by analysing some important physical quantities, such as the localization length, the ratio of the geometric mean density of states (DOS) $\rho_{\mathrm{typ}}$ to the arithmetic mean DOS $\rho_{\mathrm{ave}}$ \cite{prb_shen2012,prb_shen2013,epl_dos2003,prb_schubert2010,physrep_janssen1998,prl_dassarma2015} and the inverse participation ratio (IPR) \cite{DFS_bell1970,jpc_thouless1972,zpb_wegner1980,prl_zhang2009,physrep_janssen1998,prl_dassarma2015}.
To be specific, for a high Fermi energy [see the cyan curve in \cref{Fig1}(b)], the bulk states are localized without a magnetic field, and a $C=1$ QAH phase emerges.
However, under a strong magnetic field, localized bulk states can still form Landau bands, and the QH edge state coexists with the QAH edge state, leading to a $C=2$ phase [see \cref{Fig1}(b)].
Ultimately, we show a phase diagram of a disordered $\mathrm{MnBi_2Te_4}$ which not only relates to the current experiment \cite{Science_zhang2020}, but also indicates more coexistence states of QAH and QH waiting to be further experimentally explored.

{\textit{Effective model of ferromagnetic $\mathrm{MnBi_2Te_4}$.}}---
Due to the confinement in the $z$ direction of a three-dimensional (3D) $\mathrm{MnBi_2Te_4}$ film, a two-dimensional (2D) subband model can effectively describe the underlying physics.
Proceeding along the paradigm for the 3D to 2D crossover \cite{rmp_qixiaoliang2011}, the 2D effective model of a disordered ferromagnetic (FM) $\mathrm{MnBi_2Te_4}$ takes the following form \cite{Science_zhang2020}
\begin{equation}
  \mathcal{H}^{2 \mathrm{D}}_{\mathrm{eff}}=\left(\begin{array}{cc}
    h_{+} & \delta \\
    -\delta & h_{-}
    \end{array}\right)+H_{D},\label{eq1}
\end{equation}
where $h_{\pm}=(v_2\pm v_4)(k_x\tau_x\pm k_y\tau_y)+\left[(m_0\pm m_1)-(t_0\pm t_1)k^2\right]\tau_z$ and the coupling $\delta=it_c\tau_y$.
Here, $v_2$, $v_4$, $m_0$, $t_0$ are model parameters and $m_1$, $t_1$ are induced by FM order.
$\tau_i$ $(i=x,y,z)$ are Pauli matrices.
An external magnetic field $\Phi/\phi_0$ is included by doing the Peierls substitution \cite{prb_hofstadter1976} as a phase of the hopping term \footnote{The strength $\Phi/\phi_0$ is the magnetic flux through a unit cell in units of the flux quantum.}.
We consider the random disorder as $H_{D}=V(\mathbf{r})\tau_z$, and two parameters characterize the disorder, including the density $n$ and strength $W$ with $V(\mathbf{r})$ uniformly distributed within $[-W/2, W/2]$. Following the experiment \cite{Science_zhang2020}, the model parameters for numerical calculations are fixed as $v_2=1.41$, $v_4=1.09$, $m_0=1.025$, $m_1=0.975$, $t_0=0.5$, $t_1=1.0$, $W=4$, $n=0.2$ and $t_c=0.1$, unless otherwise specified.

To illustrate the physical picture clearly, we first simulate a six-terminal device based on \cref{eq1} under the clean limit $W=0$ and compare it with the experiment \cite{Science_zhang2020} in the Supplementary Material \cite{SM}. Evidently, the numerical results do not fit the experimental results well, especially under weak magnetic fields. However, when the disorder is involved, a detailed comparison between the numerical results and experimental data \cite{Science_zhang2020} shows a good agreement in \cref{Fig1}. The following consistency between them are obtained: 1) For a high Fermi energy $E_F=0.33$, $R_{yx}$ starts from $-h/e^2$ as a QAH phase, transits into $-h/(2e^2)$ as a C=2 Chern insulator phase [the cyan curve in \cref{Fig1}(b)]. 2) For a lower Fermi energy $E_F=0.08$, $R_{yx}$ keeps $-h/e^2$ [the purple curve in \cref{Fig1}(b)]. 3) Under a fixed magnetic field $\Phi/\phi_0=0.06$, $R_{yx}$ goes from $-h/e^2$ to $-h/(2e^2)$ as $R_{xx}$ drops to zero at the plateaux [see \cref{Fig1}(c)]. Such a good agreement confirms that our effective model can capture the underlying physics of phase transitions in ferromagnetic $\mathrm{MnBi_2Te_4}$ well.

Under the clean limit \footnote{$h_+$ decouples with $h_-$ when $t_c=0$, which is helpful for theoretical analysis. However, a small coupling $\delta$ will not change the results qualitatively.}, $h_+$ describes a QAH insulator with a gap $m_0+m_1$, while $h_-$ describes a normal insulator with a smaller gap $m_0-m_1$. Thus, when the Fermi energy $E_F$ locates between $m_0-m_1$ and $m_0+m_1$, the QAH edge state of $h_+$ is submerged by the extended bulk states from $h_-$, and thus, the Hall resistance is not quantized as shown in \cite{SM}.
As for the dirty case, the disorder tunes the extended states into localized states which is the so-called Anderson localization \cite{physrev_anderson1958}, while the QAH edge state is robust against weak disorder \cite{annualreview_liuchaoxing2016}.
Consequently, the Hall resistance shows a quantized value without a magnetic field.
In the presence of a magnetic field, the localized states can still collapse to form Landau bands, but disorder destroys them until the lowest Landau band survives. Then, the lowest Landau band forms a QH edge state which contributes the additional Hall plateau.

\begin{figure}[htbp]
  \centering
  \includegraphics[width=\columnwidth]{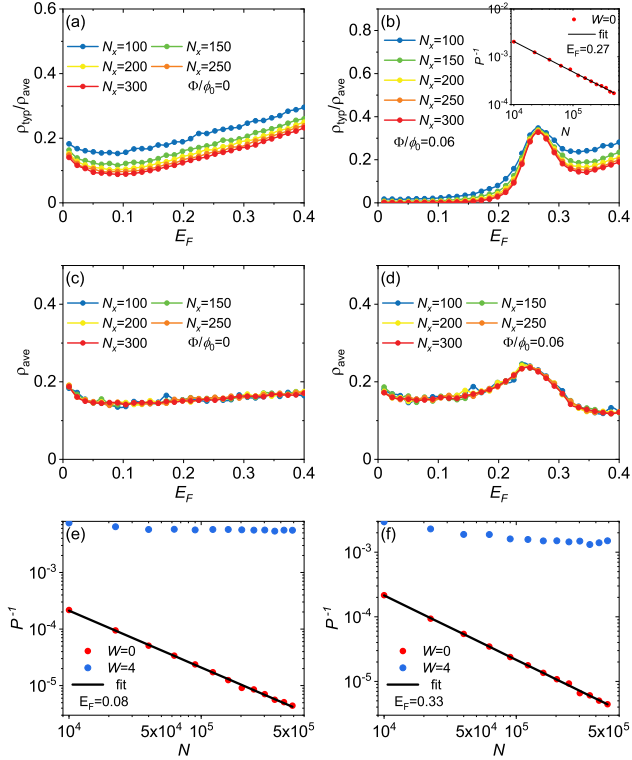}
  \caption{(color online) (a) and (b) show the ratio of the geometric mean DOS $\rho_{\mathrm{ave}}$ to the arithmetic mean DOS $\rho_{typ}$ with $\Phi/\phi_0=0$ and $\Phi/\phi_0=0.06$, respectively. The inset in (b) shows the IPR $P^{-1}$ with $E_F=0.27$. (c) and (d) show $\rho_{\mathrm{ave}}$ with $\Phi/\phi_0=0$ and $\Phi/\phi_0=0.06$, respectively. (e) and (f) display $P^{-1}$ with $E_F=0.08$ and $E_F=0.33$ in the absence of magnetic field, separately. $N=M^2$ is the total sites of the sample. The blue and red points are from numerical simulation, and the black line represents a linear fit to the data. All the data are calculated with $N_x=N_y=M$. }
  \label{Fig2}
\end{figure}

{\textit{Anderson localization induced Hall plateau.}}---
To verify the Landau bands originates from localized bulk states in the above model, we investigate the spatial extention of the eigenstates through the ratio of the geometric mean DOS $\rho_{\mathrm{typ}}$ to the arithmetic mean DOS $\rho_{\mathrm{ave}}$ \cite{physrep_janssen1998,prl_dassarma2015,prb_shen2012,prb_shen2013,epl_dos2003,prb_schubert2010} and the inverse participation ratio (IPR) \cite{DFS_bell1970,jpc_thouless1972,zpb_wegner1980,prl_zhang2009,physrep_janssen1998,prl_dassarma2015}.

$\rho_{\mathrm{ave}}$ and $\rho_{\mathrm{typ}}$ are calculated in a square sample with size $M$ under periodic boundary conditions, which are defined as \cite{prl_dassarma2015,prb_shen2012,prb_shen2013,physrep_janssen1998,epl_dos2003,prb_schubert2010}
\begin{eqnarray}
  \rho_{\mathrm{ave}}(E_F) &=& \expval{\expval{ \rho(i, E_F) }},
  \\
  \rho_{\mathrm{typ}}(E_F) &=& \exp\left[\expval{\expval{ \ln\rho(i, E_F)}}\right]
\end{eqnarray}
where $\expval{\expval{...}}$ denotes the arithmetic average over the sample sites and disorder realizations.
The local DOS $\rho(i, E_F)$ is calculated as $\rho\left(i, E_{F}\right)=\sum_{n,\alpha,\beta}\abs{\bra{i,\alpha}\ket{n,\beta}}^{2} \delta\left(E_{F}-E_{n,\beta}\right)$ where $\ket{i, \alpha}$ denotes an eigenstate at site $i$ and orbital $\alpha$, and $n$ is the index for energy level.
Here, we substitute $\eta/[\pi(x^2+\eta^2)]$ for $\delta(x)$ approximately with $\eta = 10^{-4}$ and use an exact diagonalization method \cite{prb_shen2013}.
For a extended state distributed uniformly over the sample, $\rho_\mathrm{typ}$ is almost the same as $\rho_\mathrm{ave}$, while for a localized state concentrated on certain sites, $\rho_\mathrm{typ}$ will be extremely small.
Generally, in the thermodynamic limit ($M\rightarrow\infty$), the ratio $\rho_{\mathrm{typ}}/\rho_{\mathrm{ave}}$ keeps finite for extended states, while $\rho_{\mathrm{typ}}/\rho_{\mathrm{ave}}$ approaches zero for localized states \cite{rmp_kpm2006}.

\begin{figure}[b]
  \centering
  \includegraphics[width=\columnwidth]{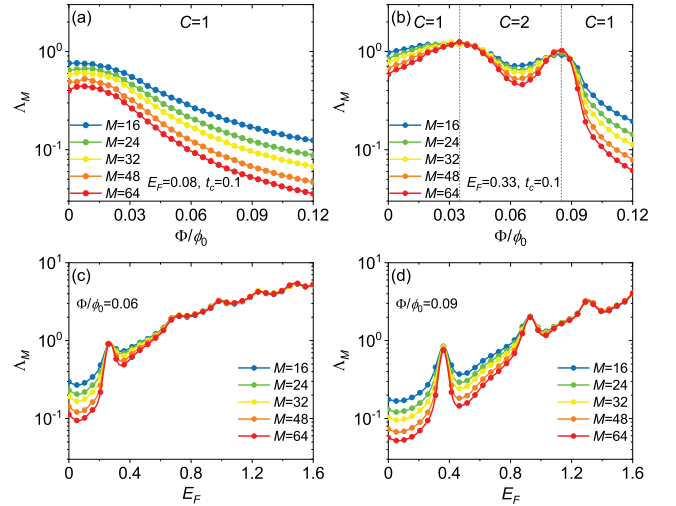}
  \caption{(color online) (a) and (b) Renormalized localization length $\Lambda_M=\lambda(M)/M$ against the magnetic field $\Phi/\phi_0$ with different Fermi energies. (c) and (d) display $\Lambda_M$ as a function of Fermi energy under different magnetic field strengths. The calculation is performed with the periodic boundary condition in the x direction.}
  \label{Fig3}
\end{figure}

In \cref{Fig2}(a)-(d), we plot $\rho_{\mathrm{typ}}/\rho_{\mathrm{ave}}$ and $\rho_{\mathrm{ave}}$ versus Fermi energy $E_F$ under different magnetic fields.
When $\Phi/\phi_0=0$, in the presence of disorder, a finite $\rho_{\mathrm{ave}}$ in \cref{Fig2}(c) and a small $\rho_{\mathrm{typ}}/\rho_{\mathrm{ave}}$ decreasing with size $M$ in \cref{Fig2}(a) indicate that the considered energy interval lies in a mobility gap rather than a bulk gap. In other words, electrons occupy the localized states.
To further reveal the effect of disorder in this case, we proceed to study the average IPR defined as \cite{zpb_wegner1980,prl_zhang2009,physrep_janssen1998,prl_dassarma2015}
\begin{equation}
  P^{-1} = \expval{\frac{\sum_{i}\abs{\psi_i}^4}{[\sum_{i}\abs{\psi_i}^2]^2}}
\end{equation}
where the wave function $\psi_i$ is evaluated at site $i$ and $E_F$, and $\expval{...}$ denotes the disorder average. As the DOS analysis, in the thermodynamic limit ($M\rightarrow\infty$), IPR scales as $P^{-1}\propto 1/N$ with $N=M^2$ for extended states, and $P^{-1}$ approaches constant for localized states \cite{jpc_thouless1972,physrep_janssen1998}.
In \cref{Fig2}(e)-(f), we investigate the effect of disorder at different $E_F$ in the absence of magnetic field.
When $W=0$, the good linear fit of the red points shows the extended nature of the wave functions for both cases.
However, when the disorder in \cref{Fig1} is included, the wave functions become localized which is manifested by the blue points with $P^{-1}$ approaching constant under large sample size $N$. It strongly shows that disorder localizes the extended states which hide the QAH edge state in the clean limit, and leads to a quantized Hall plateau with $C=1$ [see \cref{Fig1}(b)].
When the external magnetic field is applied [\cref{Fig2}(b) and 2(d)], one can identify one distinct difference from the comparison between \cref{Fig2}(b) and 2(a) that a peak arises around $E_F=0.27$. It refers to the lowest Landau band of which the extended states locate in its center. To prove the extended states, we calculate $P^{-1}$ versus total sample sites $N$ with $E_F=0.27$. As expected, the result in the inset of \cref{Fig2}(b) displays a linear relation in logarithmic coordinates. On both sides of the peak, $\rho_{\mathrm{typ}}/\rho_{\mathrm{ave}}$ is small and decreases with $M$, which indicates two different localized phases. The two phases exactly correspond to the Hall plateaux in \cref{Fig1}(c) and the extended states at $E_F=0.27$ means that the $C=2$ phase contains a QH edge state.

{\textit{Localization length.}}---
Another method to characterize the Anderson localization is to perform the finite-size scaling of localization length $\lambda(M)$.
$\lambda(M)$ is calculated at Fermi energy $E_F$ by transfer matrix method \cite{prl_mackinnon1981,zpb_mackinnon1983,repprogphys_kramer1993,njp_tomi2014,prb_zhangzhiqiang2021}.
Generally, renormalized localization length $\Lambda(M)=\lambda/M$ increases with size $M$ in a metallic phase, decreases with $M$ in a localized phase and does not depend on $M$ at the critical point.

We consider a 2D strip of length $L_y$ $(\sim 10^6)$ and width $L_x=M$, and put the finite-size scaling result in \cref{Fig3}.
First, we focus on the lower Fermi energy $E_F=0.08$, where $d\Lambda(M)/dM<0$ for any magnetic field strength [see \cref{Fig3}(a)].
According to the previous analysis, the system keeps $C=1$ as a QAH phase with a varying magnetic field. However, for $E_F=0.33$, two critical points arise [see \cref{Fig3}(b)].
With an increase in the magnetic field, the system starts from a QAH insulator with $C=1$, undergoes a phase transition, and arrives at a Chern insulator with $C=2$ which is the phase in \cref{Fig2}(b). Thus, such a $C=2$ phase is a coexistence state of QH and QAH phases.
This is consistent with the six-terminal Hall calculation in \cref{Fig1}(b).
For stronger magnetic fields, the system will come back to a QAH insulator with $C=1$ due to the shift of Landau bands.

Next, we turn to focus on $\Lambda(M)$ versus $E_F$ with fixed magnetic field [see \cref{Fig3}(c)-(d)].
When $\Phi/\phi_0=0.06$, the result in \cref{Fig3}(c) shows the distribution of Landau bands, and the peaks account for the extended states in their center.
In such a case, the lower Landau bands are more robust against disorder, whereas the higher Landau bands are so weak that $\Lambda(M)$ collapses together and peaks disappear gradually.
For a larger magnetic field, e.g. $\Phi/\phi_0=0.09$ in \cref{Fig3}(d), each Landau band moves to higher energy, and more Landau bands become stable which can contribute more Hall plateaux.

\begin{figure}[htp]
  \centering
  \includegraphics[width=\columnwidth]{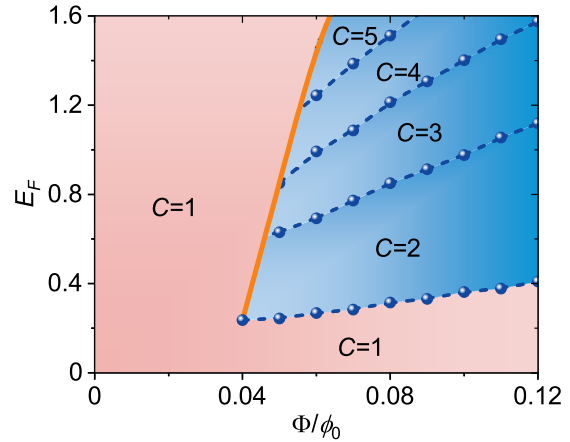}
  \caption{(color online) Phase diagram of a disordered ferromagnetic insulator $\mathrm{MnBi_2Te_4}$ in the $E_F-\Phi/\phi_0$ plane. The blue balls mark the extended states of Landau bands and the solid line indicates the phase boundary. The phase diagram is identified from the renormalized localization length $\Lambda_M$.}
  \label{Fig4}
\end{figure}

{\textit{Phase diagram and discussions.}}---
To illustrate the physical picture completely, we calculate the localization length versus Fermi energy under different magnetic fields and summarize a phase diagram in \cref{Fig4}.
Specifically, at small magnetic fields or low Fermi energies, the system is in a QAH phase with $C=1$ [the pink region in \cref{Fig4}], while for large magnetic fields or high Fermi energies, the system is in a coexistence phase of QH and QAH with $C\geqslant 2$ [the blue region in \cref{Fig4}].
For example, $E_F=0.33$ or $\Phi/\phi_0=0.06$ corresponds to the scenario observed in experiment \cite{Science_zhang2020}.
It is also remarkable to note that for a much higher Fermi energy in the phase diagram, such as $E_F=0.8$, the system will start from a $C=1$ QAH phase and go through a $C=3$ phase before arriving at the $C=2$ phase.
An evident dip of Hall resistance $R_{yx}$ towards $h/(3e^2)$ between $h/e^2$ and $h/(2e^2)$ in experiment \cite{Science_zhang2020} implies the existence of such a $C=3$ phase.
Thus, the phase diagram predicts more coexistence states of QAH and QH effect with higher Chern numbers to be verified by further experiments. To observe the higher Chern numbers, it is required to improve the quality of samples, i.e., cleaner samples, which will make it easier to realize Landau quantization under smaller magnetic fields.

{\textit{Conclusion.}}---
To summarize, we investigate a disordered $\mathrm{MnBi_2Te_4}$ under an external magnetic field and find that the exotic Hall plateaux in the experiment \cite{Science_zhang2020} originates from physics of the Anderson localization.
In the absence of a magnetic field, disorder localizes the bulk states within the QAH bulk gap and gives rise to a $C=1$ phase.
However, under an external magnetic field, the lowest Landau band, formed by those localized states, survives disorder and together with the QAH edge state, they lead to a $C=2$ phase.
Our work can help to distinguish the complex phase transitions of disordered $\mathrm{MnBi_2Te_4}$ in the presence of a magnetic field, and also stimulates further experiments to observe more coexistence states of QAH and QH.

{\textit{Acknowledgement.}}---
We thank Yuanbo Zhang for illuminating discussions.
This work is supported by the National Basic Research Program of China (Grant No. 2019YFA0308403), and the National Natural Science Foundation of China under Grants No. 11822407 and No. 11974256. C.-Z. C. is also funded by the Natural Science Foundation of Jiangsu Province under Grant No. BK20190813.

%
  

\begin{thebibliography}{44}%
  \makeatletter
  \providecommand \@ifxundefined [1]{%
   \@ifx{#1\undefined}
  }%
  \providecommand \@ifnum [1]{%
   \ifnum #1\expandafter \@firstoftwo
   \else \expandafter \@secondoftwo
   \fi
  }%
  \providecommand \@ifx [1]{%
   \ifx #1\expandafter \@firstoftwo
   \else \expandafter \@secondoftwo
   \fi
  }%
  \providecommand \natexlab [1]{#1}%
  \providecommand \enquote  [1]{``#1''}%
  \providecommand \bibnamefont  [1]{#1}%
  \providecommand \bibfnamefont [1]{#1}%
  \providecommand \citenamefont [1]{#1}%
  \providecommand \href@noop [0]{\@secondoftwo}%
  \providecommand \href [0]{\begingroup \@sanitize@url \@href}%
  \providecommand \@href[1]{\@@startlink{#1}\@@href}%
  \providecommand \@@href[1]{\endgroup#1\@@endlink}%
  \providecommand \@sanitize@url [0]{\catcode `\\12\catcode `\$12\catcode
    `\&12\catcode `\#12\catcode `\^12\catcode `\_12\catcode `\%12\relax}%
  \providecommand \@@startlink[1]{}%
  \providecommand \@@endlink[0]{}%
  \providecommand \url  [0]{\begingroup\@sanitize@url \@url }%
  \providecommand \@url [1]{\endgroup\@href {#1}{\urlprefix }}%
  \providecommand \urlprefix  [0]{URL }%
  \providecommand \Eprint [0]{\href }%
  \providecommand \doibase [0]{https://doi.org/}%
  \providecommand \selectlanguage [0]{\@gobble}%
  \providecommand \bibinfo  [0]{\@secondoftwo}%
  \providecommand \bibfield  [0]{\@secondoftwo}%
  \providecommand \translation [1]{[#1]}%
  \providecommand \BibitemOpen [0]{}%
  \providecommand \bibitemStop [0]{}%
  \providecommand \bibitemNoStop [0]{.\EOS\space}%
  \providecommand \EOS [0]{\spacefactor3000\relax}%
  \providecommand \BibitemShut  [1]{\csname bibitem#1\endcsname}%
  \let\auto@bib@innerbib\@empty
  \bibitem [{\citenamefont {Otrokov}\ \emph {et~al.}(2017)\citenamefont
    {Otrokov}, \citenamefont {Menshchikova}, \citenamefont {Vergniory},
    \citenamefont {Rusinov}, \citenamefont {Yu~Vyazovskaya}, \citenamefont
    {Koroteev}, \citenamefont {Bihlmayer}, \citenamefont {Ernst}, \citenamefont
    {Echenique}, \citenamefont {Arnau},\ and\ \citenamefont
    {Chulkov}}]{2Dmater_otrokov2017}%
    \BibitemOpen
    \bibfield  {author} {\bibinfo {author} {\bibfnamefont {M.~M.}\ \bibnamefont
    {Otrokov}}, \bibinfo {author} {\bibfnamefont {T.~V.}\ \bibnamefont
    {Menshchikova}}, \bibinfo {author} {\bibfnamefont {M.~G.}\ \bibnamefont
    {Vergniory}}, \bibinfo {author} {\bibfnamefont {I.~P.}\ \bibnamefont
    {Rusinov}}, \bibinfo {author} {\bibfnamefont {A.}~\bibnamefont
    {Yu~Vyazovskaya}}, \bibinfo {author} {\bibfnamefont {Y.~M.}\ \bibnamefont
    {Koroteev}}, \bibinfo {author} {\bibfnamefont {G.}~\bibnamefont {Bihlmayer}},
    \bibinfo {author} {\bibfnamefont {A.}~\bibnamefont {Ernst}}, \bibinfo
    {author} {\bibfnamefont {P.~M.}\ \bibnamefont {Echenique}}, \bibinfo {author}
    {\bibfnamefont {A.}~\bibnamefont {Arnau}},\ and\ \bibinfo {author}
    {\bibfnamefont {E.~V.}\ \bibnamefont {Chulkov}},\ }\bibfield  {title}
    {\bibinfo {title} {Highly-ordered wide bandgap materials for quantized
    anomalous hall and magnetoelectric effects},\ }\href
    {https://doi.org/10.1088/2053-1583/aa6bec} {\bibfield  {journal} {\bibinfo
    {journal} {2D Mater.}\ }\textbf {\bibinfo {volume} {4}},\ \bibinfo {pages}
    {025082} (\bibinfo {year} {2017})}\BibitemShut {NoStop}%
  \bibitem [{\citenamefont {Gong}\ \emph {et~al.}(2019)\citenamefont {Gong},
    \citenamefont {Guo}, \citenamefont {Li}, \citenamefont {Zhu}, \citenamefont
    {Liao}, \citenamefont {Liu}, \citenamefont {Zhang}, \citenamefont {Gu},
    \citenamefont {Tang}, \citenamefont {Feng}, \citenamefont {Zhang},
    \citenamefont {Li}, \citenamefont {Song}, \citenamefont {Wang}, \citenamefont
    {Yu}, \citenamefont {Chen}, \citenamefont {Wang}, \citenamefont {Yao},
    \citenamefont {Duan}, \citenamefont {Xu}, \citenamefont {Zhang},
    \citenamefont {Ma}, \citenamefont {Xue},\ and\ \citenamefont
    {He}}]{cpl_xue2019}%
    \BibitemOpen
    \bibfield  {author} {\bibinfo {author} {\bibfnamefont {Y.}~\bibnamefont
    {Gong}}, \bibinfo {author} {\bibfnamefont {J.}~\bibnamefont {Guo}}, \bibinfo
    {author} {\bibfnamefont {J.}~\bibnamefont {Li}}, \bibinfo {author}
    {\bibfnamefont {K.}~\bibnamefont {Zhu}}, \bibinfo {author} {\bibfnamefont
    {M.}~\bibnamefont {Liao}}, \bibinfo {author} {\bibfnamefont {X.}~\bibnamefont
    {Liu}}, \bibinfo {author} {\bibfnamefont {Q.}~\bibnamefont {Zhang}}, \bibinfo
    {author} {\bibfnamefont {L.}~\bibnamefont {Gu}}, \bibinfo {author}
    {\bibfnamefont {L.}~\bibnamefont {Tang}}, \bibinfo {author} {\bibfnamefont
    {X.}~\bibnamefont {Feng}}, \bibinfo {author} {\bibfnamefont {D.}~\bibnamefont
    {Zhang}}, \bibinfo {author} {\bibfnamefont {W.}~\bibnamefont {Li}}, \bibinfo
    {author} {\bibfnamefont {C.}~\bibnamefont {Song}}, \bibinfo {author}
    {\bibfnamefont {L.}~\bibnamefont {Wang}}, \bibinfo {author} {\bibfnamefont
    {P.}~\bibnamefont {Yu}}, \bibinfo {author} {\bibfnamefont {X.}~\bibnamefont
    {Chen}}, \bibinfo {author} {\bibfnamefont {Y.}~\bibnamefont {Wang}}, \bibinfo
    {author} {\bibfnamefont {H.}~\bibnamefont {Yao}}, \bibinfo {author}
    {\bibfnamefont {W.}~\bibnamefont {Duan}}, \bibinfo {author} {\bibfnamefont
    {Y.}~\bibnamefont {Xu}}, \bibinfo {author} {\bibfnamefont {S.-C.}\
    \bibnamefont {Zhang}}, \bibinfo {author} {\bibfnamefont {X.}~\bibnamefont
    {Ma}}, \bibinfo {author} {\bibfnamefont {Q.-K.}\ \bibnamefont {Xue}},\ and\
    \bibinfo {author} {\bibfnamefont {K.}~\bibnamefont {He}},\ }\bibfield
    {title} {\bibinfo {title} {Experimental realization of an intrinsic magnetic
    topological insulator},\ }\href
    {https://doi.org/10.1088/0256-307x/36/7/076801} {\bibfield  {journal}
    {\bibinfo  {journal} {Chinese Phys. Lett.}\ }\textbf {\bibinfo {volume}
    {36}},\ \bibinfo {pages} {076801} (\bibinfo {year} {2019})}\BibitemShut
    {NoStop}%
  \bibitem [{\citenamefont {Deng}\ \emph {et~al.}(2020)\citenamefont {Deng},
    \citenamefont {Yu}, \citenamefont {Shi}, \citenamefont {Guo}, \citenamefont
    {Xu}, \citenamefont {Wang}, \citenamefont {Chen},\ and\ \citenamefont
    {Zhang}}]{Science_zhang2020}%
    \BibitemOpen
    \bibfield  {author} {\bibinfo {author} {\bibfnamefont {Y.}~\bibnamefont
    {Deng}}, \bibinfo {author} {\bibfnamefont {Y.}~\bibnamefont {Yu}}, \bibinfo
    {author} {\bibfnamefont {M.~Z.}\ \bibnamefont {Shi}}, \bibinfo {author}
    {\bibfnamefont {Z.}~\bibnamefont {Guo}}, \bibinfo {author} {\bibfnamefont
    {Z.}~\bibnamefont {Xu}}, \bibinfo {author} {\bibfnamefont {J.}~\bibnamefont
    {Wang}}, \bibinfo {author} {\bibfnamefont {X.~H.}\ \bibnamefont {Chen}},\
    and\ \bibinfo {author} {\bibfnamefont {Y.}~\bibnamefont {Zhang}},\ }\bibfield
     {title} {\bibinfo {title} {Quantum anomalous hall effect in intrinsic
    magnetic topological insulator $\mathrm{MnBi_2Te_4}$},\ }\href
    {https://doi.org/10.1126/science.aax8156} {\bibfield  {journal} {\bibinfo
    {journal} {Science}\ }\textbf {\bibinfo {volume} {367}},\ \bibinfo {pages}
    {895} (\bibinfo {year} {2020})}\BibitemShut {NoStop}%
  \bibitem [{\citenamefont {Ge}\ \emph {et~al.}(2020)\citenamefont {Ge},
    \citenamefont {Liu}, \citenamefont {Li}, \citenamefont {Li}, \citenamefont
    {Luo}, \citenamefont {Wu}, \citenamefont {Xu},\ and\ \citenamefont
    {Wang}}]{nsr_wangjian2020}%
    \BibitemOpen
    \bibfield  {author} {\bibinfo {author} {\bibfnamefont {J.}~\bibnamefont
    {Ge}}, \bibinfo {author} {\bibfnamefont {Y.}~\bibnamefont {Liu}}, \bibinfo
    {author} {\bibfnamefont {J.}~\bibnamefont {Li}}, \bibinfo {author}
    {\bibfnamefont {H.}~\bibnamefont {Li}}, \bibinfo {author} {\bibfnamefont
    {T.}~\bibnamefont {Luo}}, \bibinfo {author} {\bibfnamefont {Y.}~\bibnamefont
    {Wu}}, \bibinfo {author} {\bibfnamefont {Y.}~\bibnamefont {Xu}},\ and\
    \bibinfo {author} {\bibfnamefont {J.}~\bibnamefont {Wang}},\ }\bibfield
    {title} {\bibinfo {title} {High-chern-number and high-temperature quantum
    hall effect without landau levels},\ }\href
    {https://doi.org/10.1093/nsr/nwaa089} {\bibfield  {journal} {\bibinfo
    {journal} {Natl. Sci. Rev.}\ }\textbf {\bibinfo {volume} {7}},\ \bibinfo
    {pages} {1280} (\bibinfo {year} {2020})}\BibitemShut {NoStop}%
  \bibitem [{\citenamefont {Liu}\ \emph {et~al.}(2020)\citenamefont {Liu},
    \citenamefont {Wang}, \citenamefont {Li}, \citenamefont {Wu}, \citenamefont
    {Li}, \citenamefont {Li}, \citenamefont {He}, \citenamefont {Xu},
    \citenamefont {Zhang},\ and\ \citenamefont {Wang}}]{natmater_wangyayu2020}%
    \BibitemOpen
    \bibfield  {author} {\bibinfo {author} {\bibfnamefont {C.}~\bibnamefont
    {Liu}}, \bibinfo {author} {\bibfnamefont {Y.}~\bibnamefont {Wang}}, \bibinfo
    {author} {\bibfnamefont {H.}~\bibnamefont {Li}}, \bibinfo {author}
    {\bibfnamefont {Y.}~\bibnamefont {Wu}}, \bibinfo {author} {\bibfnamefont
    {Y.}~\bibnamefont {Li}}, \bibinfo {author} {\bibfnamefont {J.}~\bibnamefont
    {Li}}, \bibinfo {author} {\bibfnamefont {K.}~\bibnamefont {He}}, \bibinfo
    {author} {\bibfnamefont {Y.}~\bibnamefont {Xu}}, \bibinfo {author}
    {\bibfnamefont {J.}~\bibnamefont {Zhang}},\ and\ \bibinfo {author}
    {\bibfnamefont {Y.}~\bibnamefont {Wang}},\ }\bibfield  {title} {\bibinfo
    {title} {Robust axion insulator and chern insulator phases in a
    two-dimensional antiferromagnetic topological insulator},\ }\href
    {https://doi.org/10.1038/s41563-019-0573-3} {\bibfield  {journal} {\bibinfo
    {journal} {Nat. Mater.}\ }\textbf {\bibinfo {volume} {19}},\ \bibinfo {pages}
    {522} (\bibinfo {year} {2020})}\BibitemShut {NoStop}%
  \bibitem [{\citenamefont {Zhang}\ \emph {et~al.}(2019)\citenamefont {Zhang},
    \citenamefont {Shi}, \citenamefont {Zhu}, \citenamefont {Xing}, \citenamefont
    {Zhang},\ and\ \citenamefont {Wang}}]{prl_wang2019}%
    \BibitemOpen
    \bibfield  {author} {\bibinfo {author} {\bibfnamefont {D.}~\bibnamefont
    {Zhang}}, \bibinfo {author} {\bibfnamefont {M.}~\bibnamefont {Shi}}, \bibinfo
    {author} {\bibfnamefont {T.}~\bibnamefont {Zhu}}, \bibinfo {author}
    {\bibfnamefont {D.}~\bibnamefont {Xing}}, \bibinfo {author} {\bibfnamefont
    {H.}~\bibnamefont {Zhang}},\ and\ \bibinfo {author} {\bibfnamefont
    {J.}~\bibnamefont {Wang}},\ }\bibfield  {title} {\bibinfo {title}
    {Topological axion states in the magnetic insulator
    ${\mathrm{mnbi}}_{2}{\mathrm{te}}_{4}$ with the quantized magnetoelectric
    effect},\ }\href {https://doi.org/10.1103/PhysRevLett.122.206401} {\bibfield
    {journal} {\bibinfo  {journal} {Phys. Rev. Lett.}\ }\textbf {\bibinfo
    {volume} {122}},\ \bibinfo {pages} {206401} (\bibinfo {year}
    {2019})}\BibitemShut {NoStop}%
  \bibitem [{\citenamefont {Li}\ \emph {et~al.}(2019)\citenamefont {Li},
    \citenamefont {Li}, \citenamefont {Du}, \citenamefont {Wang}, \citenamefont
    {Gu}, \citenamefont {Zhang}, \citenamefont {He}, \citenamefont {Duan},\ and\
    \citenamefont {Xu}}]{sciadv_xu2019}%
    \BibitemOpen
    \bibfield  {author} {\bibinfo {author} {\bibfnamefont {J.}~\bibnamefont
    {Li}}, \bibinfo {author} {\bibfnamefont {Y.}~\bibnamefont {Li}}, \bibinfo
    {author} {\bibfnamefont {S.}~\bibnamefont {Du}}, \bibinfo {author}
    {\bibfnamefont {Z.}~\bibnamefont {Wang}}, \bibinfo {author} {\bibfnamefont
    {B.-L.}\ \bibnamefont {Gu}}, \bibinfo {author} {\bibfnamefont {S.-C.}\
    \bibnamefont {Zhang}}, \bibinfo {author} {\bibfnamefont {K.}~\bibnamefont
    {He}}, \bibinfo {author} {\bibfnamefont {W.}~\bibnamefont {Duan}},\ and\
    \bibinfo {author} {\bibfnamefont {Y.}~\bibnamefont {Xu}},\ }\bibfield
    {title} {\bibinfo {title} {Intrinsic magnetic topological insulators in van
    der waals layered $\mathrm{MnBi_2Te_4}$-family materials},\ }\href
    {https://doi.org/10.1126/sciadv.aaw5685} {\bibfield  {journal} {\bibinfo
    {journal} {Sci. Adv.}\ }\textbf {\bibinfo {volume} {5}},\ \bibinfo {pages}
    {eaaw5685} (\bibinfo {year} {2019})}\BibitemShut {NoStop}%
  \bibitem [{\citenamefont {Otrokov}\ \emph {et~al.}(2019)\citenamefont
    {Otrokov}, \citenamefont {Klimovskikh}, \citenamefont {Bentmann},
    \citenamefont {Estyunin}, \citenamefont {Zeugner}, \citenamefont {Aliev},
    \citenamefont {Gaß}, \citenamefont {Wolter}, \citenamefont {Koroleva},
    \citenamefont {Shikin}, \citenamefont {Blanco-Rey}, \citenamefont {Hoffmann},
    \citenamefont {Rusinov}, \citenamefont {Vyazovskaya}, \citenamefont
    {Eremeev}, \citenamefont {Koroteev}, \citenamefont {Kuznetsov}, \citenamefont
    {Freyse}, \citenamefont {Sánchez-Barriga}, \citenamefont {Amiraslanov},
    \citenamefont {Babanly}, \citenamefont {Mamedov}, \citenamefont {Abdullayev},
    \citenamefont {Zverev}, \citenamefont {Alfonsov}, \citenamefont {Kataev},
    \citenamefont {Büchner}, \citenamefont {Schwier}, \citenamefont {Kumar},
    \citenamefont {Kimura}, \citenamefont {Petaccia}, \citenamefont {Di~Santo},
    \citenamefont {Vidal}, \citenamefont {Schatz}, \citenamefont {Kißner},
    \citenamefont {Ünzelmann}, \citenamefont {Min}, \citenamefont {Moser},
    \citenamefont {Peixoto}, \citenamefont {Reinert}, \citenamefont {Ernst},
    \citenamefont {Echenique}, \citenamefont {Isaeva},\ and\ \citenamefont
    {Chulkov}}]{nature_otrokov2019}%
    \BibitemOpen
    \bibfield  {author} {\bibinfo {author} {\bibfnamefont {M.~M.}\ \bibnamefont
    {Otrokov}}, \bibinfo {author} {\bibfnamefont {I.~I.}\ \bibnamefont
    {Klimovskikh}}, \bibinfo {author} {\bibfnamefont {H.}~\bibnamefont
    {Bentmann}}, \bibinfo {author} {\bibfnamefont {D.}~\bibnamefont {Estyunin}},
    \bibinfo {author} {\bibfnamefont {A.}~\bibnamefont {Zeugner}}, \bibinfo
    {author} {\bibfnamefont {Z.~S.}\ \bibnamefont {Aliev}}, \bibinfo {author}
    {\bibfnamefont {S.}~\bibnamefont {Gaß}}, \bibinfo {author} {\bibfnamefont
    {A.~U.~B.}\ \bibnamefont {Wolter}}, \bibinfo {author} {\bibfnamefont {A.~V.}\
    \bibnamefont {Koroleva}}, \bibinfo {author} {\bibfnamefont {A.~M.}\
    \bibnamefont {Shikin}}, \bibinfo {author} {\bibfnamefont {M.}~\bibnamefont
    {Blanco-Rey}}, \bibinfo {author} {\bibfnamefont {M.}~\bibnamefont
    {Hoffmann}}, \bibinfo {author} {\bibfnamefont {I.~P.}\ \bibnamefont
    {Rusinov}}, \bibinfo {author} {\bibfnamefont {A.~Y.}\ \bibnamefont
    {Vyazovskaya}}, \bibinfo {author} {\bibfnamefont {S.~V.}\ \bibnamefont
    {Eremeev}}, \bibinfo {author} {\bibfnamefont {Y.~M.}\ \bibnamefont
    {Koroteev}}, \bibinfo {author} {\bibfnamefont {V.~M.}\ \bibnamefont
    {Kuznetsov}}, \bibinfo {author} {\bibfnamefont {F.}~\bibnamefont {Freyse}},
    \bibinfo {author} {\bibfnamefont {J.}~\bibnamefont {Sánchez-Barriga}},
    \bibinfo {author} {\bibfnamefont {I.~R.}\ \bibnamefont {Amiraslanov}},
    \bibinfo {author} {\bibfnamefont {M.~B.}\ \bibnamefont {Babanly}}, \bibinfo
    {author} {\bibfnamefont {N.~T.}\ \bibnamefont {Mamedov}}, \bibinfo {author}
    {\bibfnamefont {N.~A.}\ \bibnamefont {Abdullayev}}, \bibinfo {author}
    {\bibfnamefont {V.~N.}\ \bibnamefont {Zverev}}, \bibinfo {author}
    {\bibfnamefont {A.}~\bibnamefont {Alfonsov}}, \bibinfo {author}
    {\bibfnamefont {V.}~\bibnamefont {Kataev}}, \bibinfo {author} {\bibfnamefont
    {B.}~\bibnamefont {Büchner}}, \bibinfo {author} {\bibfnamefont {E.~F.}\
    \bibnamefont {Schwier}}, \bibinfo {author} {\bibfnamefont {S.}~\bibnamefont
    {Kumar}}, \bibinfo {author} {\bibfnamefont {A.}~\bibnamefont {Kimura}},
    \bibinfo {author} {\bibfnamefont {L.}~\bibnamefont {Petaccia}}, \bibinfo
    {author} {\bibfnamefont {G.}~\bibnamefont {Di~Santo}}, \bibinfo {author}
    {\bibfnamefont {R.~C.}\ \bibnamefont {Vidal}}, \bibinfo {author}
    {\bibfnamefont {S.}~\bibnamefont {Schatz}}, \bibinfo {author} {\bibfnamefont
    {K.}~\bibnamefont {Kißner}}, \bibinfo {author} {\bibfnamefont
    {M.}~\bibnamefont {Ünzelmann}}, \bibinfo {author} {\bibfnamefont {C.~H.}\
    \bibnamefont {Min}}, \bibinfo {author} {\bibfnamefont {S.}~\bibnamefont
    {Moser}}, \bibinfo {author} {\bibfnamefont {T.~R.~F.}\ \bibnamefont
    {Peixoto}}, \bibinfo {author} {\bibfnamefont {F.}~\bibnamefont {Reinert}},
    \bibinfo {author} {\bibfnamefont {A.}~\bibnamefont {Ernst}}, \bibinfo
    {author} {\bibfnamefont {P.~M.}\ \bibnamefont {Echenique}}, \bibinfo {author}
    {\bibfnamefont {A.}~\bibnamefont {Isaeva}},\ and\ \bibinfo {author}
    {\bibfnamefont {E.~V.}\ \bibnamefont {Chulkov}},\ }\bibfield  {title}
    {\bibinfo {title} {Prediction and observation of an antiferromagnetic
    topological insulator},\ }\href {https://doi.org/10.1038/s41586-019-1840-9}
    {\bibfield  {journal} {\bibinfo  {journal} {Nature}\ }\textbf {\bibinfo
    {volume} {576}},\ \bibinfo {pages} {416} (\bibinfo {year}
    {2019})}\BibitemShut {NoStop}%
  \bibitem [{\citenamefont {Zhang}\ \emph {et~al.}(2020)\citenamefont {Zhang},
    \citenamefont {Wu},\ and\ \citenamefont {Das~Sarma}}]{prl_dassarma2020}%
    \BibitemOpen
    \bibfield  {author} {\bibinfo {author} {\bibfnamefont {R.-X.}\ \bibnamefont
    {Zhang}}, \bibinfo {author} {\bibfnamefont {F.}~\bibnamefont {Wu}},\ and\
    \bibinfo {author} {\bibfnamefont {S.}~\bibnamefont {Das~Sarma}},\ }\bibfield
    {title} {\bibinfo {title} {M\"obius insulator and higher-order topology in
    ${\mathrm{mnbi}}_{2n}{\mathrm{te}}_{3n+1}$},\ }\href
    {https://doi.org/10.1103/PhysRevLett.124.136407} {\bibfield  {journal}
    {\bibinfo  {journal} {Phys. Rev. Lett.}\ }\textbf {\bibinfo {volume} {124}},\
    \bibinfo {pages} {136407} (\bibinfo {year} {2020})}\BibitemShut {NoStop}%
  \bibitem [{\citenamefont {Hao}\ \emph {et~al.}(2019)\citenamefont {Hao},
    \citenamefont {Liu}, \citenamefont {Feng}, \citenamefont {Ma}, \citenamefont
    {Schwier}, \citenamefont {Arita}, \citenamefont {Kumar}, \citenamefont {Hu},
    \citenamefont {Lu}, \citenamefont {Zeng}, \citenamefont {Wang}, \citenamefont
    {Hao}, \citenamefont {Sun}, \citenamefont {Zhang}, \citenamefont {Mei},
    \citenamefont {Ni}, \citenamefont {Wu}, \citenamefont {Shimada},
    \citenamefont {Chen}, \citenamefont {Liu},\ and\ \citenamefont
    {Liu}}]{prx_hao2019}%
    \BibitemOpen
    \bibfield  {author} {\bibinfo {author} {\bibfnamefont {Y.-J.}\ \bibnamefont
    {Hao}}, \bibinfo {author} {\bibfnamefont {P.}~\bibnamefont {Liu}}, \bibinfo
    {author} {\bibfnamefont {Y.}~\bibnamefont {Feng}}, \bibinfo {author}
    {\bibfnamefont {X.-M.}\ \bibnamefont {Ma}}, \bibinfo {author} {\bibfnamefont
    {E.~F.}\ \bibnamefont {Schwier}}, \bibinfo {author} {\bibfnamefont
    {M.}~\bibnamefont {Arita}}, \bibinfo {author} {\bibfnamefont
    {S.}~\bibnamefont {Kumar}}, \bibinfo {author} {\bibfnamefont
    {C.}~\bibnamefont {Hu}}, \bibinfo {author} {\bibfnamefont {R.}~\bibnamefont
    {Lu}}, \bibinfo {author} {\bibfnamefont {M.}~\bibnamefont {Zeng}}, \bibinfo
    {author} {\bibfnamefont {Y.}~\bibnamefont {Wang}}, \bibinfo {author}
    {\bibfnamefont {Z.}~\bibnamefont {Hao}}, \bibinfo {author} {\bibfnamefont
    {H.-Y.}\ \bibnamefont {Sun}}, \bibinfo {author} {\bibfnamefont
    {K.}~\bibnamefont {Zhang}}, \bibinfo {author} {\bibfnamefont
    {J.}~\bibnamefont {Mei}}, \bibinfo {author} {\bibfnamefont {N.}~\bibnamefont
    {Ni}}, \bibinfo {author} {\bibfnamefont {L.}~\bibnamefont {Wu}}, \bibinfo
    {author} {\bibfnamefont {K.}~\bibnamefont {Shimada}}, \bibinfo {author}
    {\bibfnamefont {C.}~\bibnamefont {Chen}}, \bibinfo {author} {\bibfnamefont
    {Q.}~\bibnamefont {Liu}},\ and\ \bibinfo {author} {\bibfnamefont
    {C.}~\bibnamefont {Liu}},\ }\bibfield  {title} {\bibinfo {title} {Gapless
    surface dirac cone in antiferromagnetic topological insulator
    ${\mathrm{mnbi}}_{2}{\mathrm{te}}_{4}$},\ }\href
    {https://doi.org/10.1103/PhysRevX.9.041038} {\bibfield  {journal} {\bibinfo
    {journal} {Phys. Rev. X}\ }\textbf {\bibinfo {volume} {9}},\ \bibinfo {pages}
    {041038} (\bibinfo {year} {2019})}\BibitemShut {NoStop}%
  \bibitem [{\citenamefont {Liu}\ \emph {et~al.}(2008)\citenamefont {Liu},
    \citenamefont {Qi}, \citenamefont {Dai}, \citenamefont {Fang},\ and\
    \citenamefont {Zhang}}]{prl_qah2008}%
    \BibitemOpen
    \bibfield  {author} {\bibinfo {author} {\bibfnamefont {C.-X.}\ \bibnamefont
    {Liu}}, \bibinfo {author} {\bibfnamefont {X.-L.}\ \bibnamefont {Qi}},
    \bibinfo {author} {\bibfnamefont {X.}~\bibnamefont {Dai}}, \bibinfo {author}
    {\bibfnamefont {Z.}~\bibnamefont {Fang}},\ and\ \bibinfo {author}
    {\bibfnamefont {S.-C.}\ \bibnamefont {Zhang}},\ }\bibfield  {title} {\bibinfo
    {title} {Quantum anomalous hall effect in
    ${\mathrm{hg}}_{1\ensuremath{-}y}{\mathrm{mn}}_{y}\mathrm{Te}$ quantum
    wells},\ }\href {https://doi.org/10.1103/PhysRevLett.101.146802} {\bibfield
    {journal} {\bibinfo  {journal} {Phys. Rev. Lett.}\ }\textbf {\bibinfo
    {volume} {101}},\ \bibinfo {pages} {146802} (\bibinfo {year}
    {2008})}\BibitemShut {NoStop}%
  \bibitem [{\citenamefont {Yu}\ \emph {et~al.}(2010)\citenamefont {Yu},
    \citenamefont {Zhang}, \citenamefont {Zhang}, \citenamefont {Zhang},
    \citenamefont {Dai},\ and\ \citenamefont {Fang}}]{science_qah2010}%
    \BibitemOpen
    \bibfield  {author} {\bibinfo {author} {\bibfnamefont {R.}~\bibnamefont
    {Yu}}, \bibinfo {author} {\bibfnamefont {W.}~\bibnamefont {Zhang}}, \bibinfo
    {author} {\bibfnamefont {H.-J.}\ \bibnamefont {Zhang}}, \bibinfo {author}
    {\bibfnamefont {S.-C.}\ \bibnamefont {Zhang}}, \bibinfo {author}
    {\bibfnamefont {X.}~\bibnamefont {Dai}},\ and\ \bibinfo {author}
    {\bibfnamefont {Z.}~\bibnamefont {Fang}},\ }\bibfield  {title} {\bibinfo
    {title} {Quantized anomalous hall effect in magnetic topological
    insulators},\ }\href {https://doi.org/10.1126/science.1187485 %J Science}
    {\bibfield  {journal} {\bibinfo  {journal} {Science}\ }\textbf {\bibinfo
    {volume} {329}},\ \bibinfo {pages} {61} (\bibinfo {year} {2010})}\BibitemShut
    {NoStop}%
  \bibitem [{\citenamefont {Chang}\ \emph {et~al.}(2013)\citenamefont {Chang},
    \citenamefont {Zhang}, \citenamefont {Feng}, \citenamefont {Shen},
    \citenamefont {Zhang}, \citenamefont {Guo}, \citenamefont {Li}, \citenamefont
    {Ou}, \citenamefont {Wei}, \citenamefont {Wang}, \citenamefont {Ji},
    \citenamefont {Feng}, \citenamefont {Ji}, \citenamefont {Chen}, \citenamefont
    {Jia}, \citenamefont {Dai}, \citenamefont {Fang}, \citenamefont {Zhang},
    \citenamefont {He}, \citenamefont {Wang}, \citenamefont {Lu}, \citenamefont
    {Ma},\ and\ \citenamefont {Xue}}]{science_qah2013}%
    \BibitemOpen
    \bibfield  {author} {\bibinfo {author} {\bibfnamefont {C.-Z.}\ \bibnamefont
    {Chang}}, \bibinfo {author} {\bibfnamefont {J.}~\bibnamefont {Zhang}},
    \bibinfo {author} {\bibfnamefont {X.}~\bibnamefont {Feng}}, \bibinfo {author}
    {\bibfnamefont {J.}~\bibnamefont {Shen}}, \bibinfo {author} {\bibfnamefont
    {Z.}~\bibnamefont {Zhang}}, \bibinfo {author} {\bibfnamefont
    {M.}~\bibnamefont {Guo}}, \bibinfo {author} {\bibfnamefont {K.}~\bibnamefont
    {Li}}, \bibinfo {author} {\bibfnamefont {Y.}~\bibnamefont {Ou}}, \bibinfo
    {author} {\bibfnamefont {P.}~\bibnamefont {Wei}}, \bibinfo {author}
    {\bibfnamefont {L.-L.}\ \bibnamefont {Wang}}, \bibinfo {author}
    {\bibfnamefont {Z.-Q.}\ \bibnamefont {Ji}}, \bibinfo {author} {\bibfnamefont
    {Y.}~\bibnamefont {Feng}}, \bibinfo {author} {\bibfnamefont {S.}~\bibnamefont
    {Ji}}, \bibinfo {author} {\bibfnamefont {X.}~\bibnamefont {Chen}}, \bibinfo
    {author} {\bibfnamefont {J.}~\bibnamefont {Jia}}, \bibinfo {author}
    {\bibfnamefont {X.}~\bibnamefont {Dai}}, \bibinfo {author} {\bibfnamefont
    {Z.}~\bibnamefont {Fang}}, \bibinfo {author} {\bibfnamefont {S.-C.}\
    \bibnamefont {Zhang}}, \bibinfo {author} {\bibfnamefont {K.}~\bibnamefont
    {He}}, \bibinfo {author} {\bibfnamefont {Y.}~\bibnamefont {Wang}}, \bibinfo
    {author} {\bibfnamefont {L.}~\bibnamefont {Lu}}, \bibinfo {author}
    {\bibfnamefont {X.-C.}\ \bibnamefont {Ma}},\ and\ \bibinfo {author}
    {\bibfnamefont {Q.-K.}\ \bibnamefont {Xue}},\ }\bibfield  {title} {\bibinfo
    {title} {Experimental observation of the quantum anomalous hall effect in a
    magnetic topological insulator},\ }\href
    {https://doi.org/10.1126/science.1234414} {\bibfield  {journal} {\bibinfo
    {journal} {Science}\ }\textbf {\bibinfo {volume} {340}},\ \bibinfo {pages}
    {167} (\bibinfo {year} {2013})}\BibitemShut {NoStop}%
  \bibitem [{\citenamefont {Liu}\ \emph {et~al.}(2016{\natexlab{a}})\citenamefont
    {Liu}, \citenamefont {Zhang},\ and\ \citenamefont {Qi}}]{arp_qah2016}%
    \BibitemOpen
    \bibfield  {author} {\bibinfo {author} {\bibfnamefont {C.-X.}\ \bibnamefont
    {Liu}}, \bibinfo {author} {\bibfnamefont {S.-C.}\ \bibnamefont {Zhang}},\
    and\ \bibinfo {author} {\bibfnamefont {X.-L.}\ \bibnamefont {Qi}},\
    }\bibfield  {title} {\bibinfo {title} {The quantum anomalous hall effect:
    Theory and experiment},\ }\href
    {https://doi.org/10.1146/annurev-conmatphys-031115-011417} {\bibfield
    {journal} {\bibinfo  {journal} {Annu. Rev. Condens. Matter Phys.}\ }\textbf
    {\bibinfo {volume} {7}},\ \bibinfo {pages} {301} (\bibinfo {year}
    {2016}{\natexlab{a}})}\BibitemShut {NoStop}%
  \bibitem [{\citenamefont {Anderson}(1958)}]{physrev_anderson1958}%
    \BibitemOpen
    \bibfield  {author} {\bibinfo {author} {\bibfnamefont {P.~W.}\ \bibnamefont
    {Anderson}},\ }\bibfield  {title} {\bibinfo {title} {Absence of diffusion in
    certain random lattices},\ }\href {https://doi.org/10.1103/PhysRev.109.1492}
    {\bibfield  {journal} {\bibinfo  {journal} {Phys. Rev.}\ }\textbf {\bibinfo
    {volume} {109}},\ \bibinfo {pages} {1492} (\bibinfo {year}
    {1958})}\BibitemShut {NoStop}%
  \bibitem [{\citenamefont {Onoda}\ \emph {et~al.}(2007)\citenamefont {Onoda},
    \citenamefont {Avishai},\ and\ \citenamefont {Nagaosa}}]{prl_nagaosa2007}%
    \BibitemOpen
    \bibfield  {author} {\bibinfo {author} {\bibfnamefont {M.}~\bibnamefont
    {Onoda}}, \bibinfo {author} {\bibfnamefont {Y.}~\bibnamefont {Avishai}},\
    and\ \bibinfo {author} {\bibfnamefont {N.}~\bibnamefont {Nagaosa}},\
    }\bibfield  {title} {\bibinfo {title} {Localization in a quantum spin hall
    system},\ }\href {https://doi.org/10.1103/PhysRevLett.98.076802} {\bibfield
    {journal} {\bibinfo  {journal} {Phys. Rev. Lett.}\ }\textbf {\bibinfo
    {volume} {98}},\ \bibinfo {pages} {076802} (\bibinfo {year}
    {2007})}\BibitemShut {NoStop}%
  \bibitem [{\citenamefont {Hikami}\ \emph {et~al.}(1980)\citenamefont {Hikami},
    \citenamefont {Larkin},\ and\ \citenamefont
    {Nagaoka}}]{progtheorphys_hikami1980}%
    \BibitemOpen
    \bibfield  {author} {\bibinfo {author} {\bibfnamefont {S.}~\bibnamefont
    {Hikami}}, \bibinfo {author} {\bibfnamefont {A.~I.}\ \bibnamefont {Larkin}},\
    and\ \bibinfo {author} {\bibfnamefont {Y.}~\bibnamefont {Nagaoka}},\
    }\bibfield  {title} {\bibinfo {title} {Spin-orbit interaction and
    magnetoresistance in the two dimensional random system},\ }\href
    {https://doi.org/10.1143/PTP.63.707 %J Progress of Theoretical Physics}
    {\bibfield  {journal} {\bibinfo  {journal} {Prog. Theor. Phys.}\ }\textbf
    {\bibinfo {volume} {63}},\ \bibinfo {pages} {707} (\bibinfo {year}
    {1980})}\BibitemShut {NoStop}%
  \bibitem [{\citenamefont {Ando}(1989)}]{prb_ando1989}%
    \BibitemOpen
    \bibfield  {author} {\bibinfo {author} {\bibfnamefont {T.}~\bibnamefont
    {Ando}},\ }\bibfield  {title} {\bibinfo {title} {Numerical study of symmetry
    effects on localization in two dimensions},\ }\href
    {https://doi.org/10.1103/PhysRevB.40.5325} {\bibfield  {journal} {\bibinfo
    {journal} {Phys.l Rev. B}\ }\textbf {\bibinfo {volume} {40}},\ \bibinfo
    {pages} {5325} (\bibinfo {year} {1989})}\BibitemShut {NoStop}%
  \bibitem [{\citenamefont {Azbel’}(1992)}]{prb_azbel1992}%
    \BibitemOpen
    \bibfield  {author} {\bibinfo {author} {\bibfnamefont {M.~Y.}\ \bibnamefont
    {Azbel’}},\ }\bibfield  {title} {\bibinfo {title} {Quantum particle in a
    random potential: Exact solution and its implications},\ }\href
    {https://doi.org/10.1103/PhysRevB.45.4208} {\bibfield  {journal} {\bibinfo
    {journal} {Phys. Rev. B}\ }\textbf {\bibinfo {volume} {45}},\ \bibinfo
    {pages} {4208} (\bibinfo {year} {1992})}\BibitemShut {NoStop}%
  \bibitem [{\citenamefont {MacKinnon}(1985)}]{zpb_mackinnon1985}%
    \BibitemOpen
    \bibfield  {author} {\bibinfo {author} {\bibfnamefont {A.}~\bibnamefont
    {MacKinnon}},\ }\bibfield  {title} {\bibinfo {title} {The calculation of
    transport properties and density of states of disordered solids},\ }\href
    {https://doi.org/10.1007/bf01328846} {\bibfield  {journal} {\bibinfo
    {journal} {Z. Phys. B}\ }\textbf {\bibinfo {volume} {59}},\ \bibinfo {pages}
    {385} (\bibinfo {year} {1985})}\BibitemShut {NoStop}%
  \bibitem [{\citenamefont {Jauho}\ \emph {et~al.}(1994)\citenamefont {Jauho},
    \citenamefont {Wingreen},\ and\ \citenamefont {Meir}}]{prb_jauho1994}%
    \BibitemOpen
    \bibfield  {author} {\bibinfo {author} {\bibfnamefont {A.-P.}\ \bibnamefont
    {Jauho}}, \bibinfo {author} {\bibfnamefont {N.~S.}\ \bibnamefont
    {Wingreen}},\ and\ \bibinfo {author} {\bibfnamefont {Y.}~\bibnamefont
    {Meir}},\ }\bibfield  {title} {\bibinfo {title} {Time-dependent transport in
    interacting and noninteracting resonant-tunneling systems},\ }\href
    {https://doi.org/10.1103/PhysRevB.50.5528} {\bibfield  {journal} {\bibinfo
    {journal} {Phys. Rev. B}\ }\textbf {\bibinfo {volume} {50}},\ \bibinfo
    {pages} {5528} (\bibinfo {year} {1994})}\BibitemShut {NoStop}%
  \bibitem [{\citenamefont {Camsari}\ \emph {et~al.}(2020)\citenamefont
    {Camsari}, \citenamefont {Chowdhury},\ and\ \citenamefont
    {Datta}}]{arxiv_datta2020}%
    \BibitemOpen
    \bibfield  {author} {\bibinfo {author} {\bibfnamefont {K.~Y.}\ \bibnamefont
    {Camsari}}, \bibinfo {author} {\bibfnamefont {S.}~\bibnamefont {Chowdhury}},\
    and\ \bibinfo {author} {\bibfnamefont {S.}~\bibnamefont {Datta}},\ }\bibfield
     {title} {\bibinfo {title} {The non-equilibrium green function (negf)
    method},\ }\href {https://arxiv.org/abs/2008.01275} {\bibfield  {journal}
    {\bibinfo  {journal} {arXiv e-prints}\ ,\ \bibinfo {pages}
    {arXiv:2008.01275}} (\bibinfo {year} {2020})}\BibitemShut {NoStop}%
  \bibitem [{\citenamefont {Zhang}\ \emph {et~al.}(2012)\citenamefont {Zhang},
    \citenamefont {Chu}, \citenamefont {Zhang},\ and\ \citenamefont
    {Shen}}]{prb_shen2012}%
    \BibitemOpen
    \bibfield  {author} {\bibinfo {author} {\bibfnamefont {Y.-Y.}\ \bibnamefont
    {Zhang}}, \bibinfo {author} {\bibfnamefont {R.-L.}\ \bibnamefont {Chu}},
    \bibinfo {author} {\bibfnamefont {F.-C.}\ \bibnamefont {Zhang}},\ and\
    \bibinfo {author} {\bibfnamefont {S.-Q.}\ \bibnamefont {Shen}},\ }\bibfield
    {title} {\bibinfo {title} {Localization and mobility gap in the topological
    anderson insulator},\ }\href {https://doi.org/10.1103/PhysRevB.85.035107}
    {\bibfield  {journal} {\bibinfo  {journal} {Phys. Rev. B}\ }\textbf {\bibinfo
    {volume} {85}},\ \bibinfo {pages} {035107} (\bibinfo {year}
    {2012})}\BibitemShut {NoStop}%
  \bibitem [{\citenamefont {Zhang}\ and\ \citenamefont
    {Shen}(2013)}]{prb_shen2013}%
    \BibitemOpen
    \bibfield  {author} {\bibinfo {author} {\bibfnamefont {Y.-Y.}\ \bibnamefont
    {Zhang}}\ and\ \bibinfo {author} {\bibfnamefont {S.-Q.}\ \bibnamefont
    {Shen}},\ }\bibfield  {title} {\bibinfo {title} {Algebraic and geometric mean
    density of states in topological anderson insulators},\ }\href
    {https://doi.org/10.1103/PhysRevB.88.195145} {\bibfield  {journal} {\bibinfo
    {journal} {Phys. Rev. B}\ }\textbf {\bibinfo {volume} {88}},\ \bibinfo
    {pages} {195145} (\bibinfo {year} {2013})}\BibitemShut {NoStop}%
  \bibitem [{\citenamefont {Dobrosavljević}\ \emph {et~al.}(2003)\citenamefont
    {Dobrosavljević}, \citenamefont {Pastor},\ and\ \citenamefont
    {Nikolić}}]{epl_dos2003}%
    \BibitemOpen
    \bibfield  {author} {\bibinfo {author} {\bibfnamefont {V.}~\bibnamefont
    {Dobrosavljević}}, \bibinfo {author} {\bibfnamefont {A.~A.}\ \bibnamefont
    {Pastor}},\ and\ \bibinfo {author} {\bibfnamefont {B.~K.}\ \bibnamefont
    {Nikolić}},\ }\bibfield  {title} {\bibinfo {title} {Typical medium theory of
    anderson localization: A local order parameter approach to strong-disorder
    effects},\ }\href {https://doi.org/10.1209/epl/i2003-00364-5} {\bibfield
    {journal} {\bibinfo  {journal} {Europhys. Lett.}\ }\textbf {\bibinfo {volume}
    {62}},\ \bibinfo {pages} {76} (\bibinfo {year} {2003})}\BibitemShut {NoStop}%
  \bibitem [{\citenamefont {Schubert}\ \emph {et~al.}(2010)\citenamefont
    {Schubert}, \citenamefont {Schleede}, \citenamefont {Byczuk}, \citenamefont
    {Fehske},\ and\ \citenamefont {Vollhardt}}]{prb_schubert2010}%
    \BibitemOpen
    \bibfield  {author} {\bibinfo {author} {\bibfnamefont {G.}~\bibnamefont
    {Schubert}}, \bibinfo {author} {\bibfnamefont {J.}~\bibnamefont {Schleede}},
    \bibinfo {author} {\bibfnamefont {K.}~\bibnamefont {Byczuk}}, \bibinfo
    {author} {\bibfnamefont {H.}~\bibnamefont {Fehske}},\ and\ \bibinfo {author}
    {\bibfnamefont {D.}~\bibnamefont {Vollhardt}},\ }\bibfield  {title} {\bibinfo
    {title} {Distribution of the local density of states as a criterion for
    anderson localization: Numerically exact results for various lattices in two
    and three dimensions},\ }\href {https://doi.org/10.1103/PhysRevB.81.155106}
    {\bibfield  {journal} {\bibinfo  {journal} {Phys. Rev. B}\ }\textbf {\bibinfo
    {volume} {81}},\ \bibinfo {pages} {155106} (\bibinfo {year}
    {2010})}\BibitemShut {NoStop}%
  \bibitem [{\citenamefont {Janssen}(1998)}]{physrep_janssen1998}%
    \BibitemOpen
    \bibfield  {author} {\bibinfo {author} {\bibfnamefont {M.}~\bibnamefont
    {Janssen}},\ }\bibfield  {title} {\bibinfo {title} {Statistics and scaling in
    disordered mesoscopic electron systems},\ }\href
    {https://doi.org/https://doi.org/10.1016/S0370-1573(97)00050-1} {\bibfield
    {journal} {\bibinfo  {journal} {Phys. Rep.}\ }\textbf {\bibinfo {volume}
    {295}},\ \bibinfo {pages} {1} (\bibinfo {year} {1998})}\BibitemShut {NoStop}%
  \bibitem [{\citenamefont {Pixley}\ \emph {et~al.}(2015)\citenamefont {Pixley},
    \citenamefont {Goswami},\ and\ \citenamefont {Das~Sarma}}]{prl_dassarma2015}%
    \BibitemOpen
    \bibfield  {author} {\bibinfo {author} {\bibfnamefont {J.~H.}\ \bibnamefont
    {Pixley}}, \bibinfo {author} {\bibfnamefont {P.}~\bibnamefont {Goswami}},\
    and\ \bibinfo {author} {\bibfnamefont {S.}~\bibnamefont {Das~Sarma}},\
    }\bibfield  {title} {\bibinfo {title} {Anderson localization and the quantum
    phase diagram of three dimensional disordered dirac semimetals},\ }\href
    {https://doi.org/10.1103/PhysRevLett.115.076601} {\bibfield  {journal}
    {\bibinfo  {journal} {Phys. Rev. Lett.}\ }\textbf {\bibinfo {volume} {115}},\
    \bibinfo {pages} {076601} (\bibinfo {year} {2015})}\BibitemShut {NoStop}%
  \bibitem [{\citenamefont {Bell}\ and\ \citenamefont
    {Dean}(1970)}]{DFS_bell1970}%
    \BibitemOpen
    \bibfield  {author} {\bibinfo {author} {\bibfnamefont {R.~J.}\ \bibnamefont
    {Bell}}\ and\ \bibinfo {author} {\bibfnamefont {P.}~\bibnamefont {Dean}},\
    }\bibfield  {title} {\bibinfo {title} {Atomic vibrations in vitreous
    silica},\ }\href {https://doi.org/10.1039/DF9705000055} {\bibfield  {journal}
    {\bibinfo  {journal} {Discuss. Faraday Soc.}\ }\textbf {\bibinfo {volume}
    {50}},\ \bibinfo {pages} {55} (\bibinfo {year} {1970})}\BibitemShut {NoStop}%
  \bibitem [{\citenamefont {Edwards}\ and\ \citenamefont
    {Thouless}(1972)}]{jpc_thouless1972}%
    \BibitemOpen
    \bibfield  {author} {\bibinfo {author} {\bibfnamefont {J.~T.}\ \bibnamefont
    {Edwards}}\ and\ \bibinfo {author} {\bibfnamefont {D.~J.}\ \bibnamefont
    {Thouless}},\ }\bibfield  {title} {\bibinfo {title} {Numerical studies of
    localization in disordered systems},\ }\href
    {https://doi.org/10.1088/0022-3719/5/8/007} {\bibfield  {journal} {\bibinfo
    {journal} {J. Phy. C}\ }\textbf {\bibinfo {volume} {5}},\ \bibinfo {pages}
    {807} (\bibinfo {year} {1972})}\BibitemShut {NoStop}%
  \bibitem [{\citenamefont {Wegner}(1980)}]{zpb_wegner1980}%
    \BibitemOpen
    \bibfield  {author} {\bibinfo {author} {\bibfnamefont {F.}~\bibnamefont
    {Wegner}},\ }\bibfield  {title} {\bibinfo {title} {Inverse participation
    ratio in $2+\epsilon$ dimensions},\ }\href
    {https://doi.org/10.1007/BF01325284} {\bibfield  {journal} {\bibinfo
    {journal} {Z. Phys. B}\ }\textbf {\bibinfo {volume} {36}},\ \bibinfo {pages}
    {209} (\bibinfo {year} {1980})}\BibitemShut {NoStop}%
  \bibitem [{\citenamefont {Zhang}\ \emph {et~al.}(2009)\citenamefont {Zhang},
    \citenamefont {Hu}, \citenamefont {Bernevig}, \citenamefont {Wang},
    \citenamefont {Xie},\ and\ \citenamefont {Liu}}]{prl_zhang2009}%
    \BibitemOpen
    \bibfield  {author} {\bibinfo {author} {\bibfnamefont {Y.-Y.}\ \bibnamefont
    {Zhang}}, \bibinfo {author} {\bibfnamefont {J.}~\bibnamefont {Hu}}, \bibinfo
    {author} {\bibfnamefont {B.~A.}\ \bibnamefont {Bernevig}}, \bibinfo {author}
    {\bibfnamefont {X.~R.}\ \bibnamefont {Wang}}, \bibinfo {author}
    {\bibfnamefont {X.~C.}\ \bibnamefont {Xie}},\ and\ \bibinfo {author}
    {\bibfnamefont {W.~M.}\ \bibnamefont {Liu}},\ }\bibfield  {title} {\bibinfo
    {title} {Localization and the kosterlitz-thouless transition in disordered
    graphene},\ }\href {https://doi.org/10.1103/PhysRevLett.102.106401}
    {\bibfield  {journal} {\bibinfo  {journal} {Phys. Rev. Lett.}\ }\textbf
    {\bibinfo {volume} {102}},\ \bibinfo {pages} {106401} (\bibinfo {year}
    {2009})}\BibitemShut {NoStop}%
  \bibitem [{\citenamefont {Qi}\ and\ \citenamefont
    {Zhang}(2011)}]{rmp_qixiaoliang2011}%
    \BibitemOpen
    \bibfield  {author} {\bibinfo {author} {\bibfnamefont {X.-L.}\ \bibnamefont
    {Qi}}\ and\ \bibinfo {author} {\bibfnamefont {S.-C.}\ \bibnamefont {Zhang}},\
    }\bibfield  {title} {\bibinfo {title} {Topological insulators and
    superconductors},\ }\href {https://doi.org/10.1103/RevModPhys.83.1057}
    {\bibfield  {journal} {\bibinfo  {journal} {Rev. Mod. Phys.}\ }\textbf
    {\bibinfo {volume} {83}},\ \bibinfo {pages} {1057} (\bibinfo {year}
    {2011})}\BibitemShut {NoStop}%
  \bibitem [{\citenamefont {Hofstadter}(1976)}]{prb_hofstadter1976}%
    \BibitemOpen
    \bibfield  {author} {\bibinfo {author} {\bibfnamefont {D.~R.}\ \bibnamefont
    {Hofstadter}},\ }\bibfield  {title} {\bibinfo {title} {Energy levels and wave
    functions of bloch electrons in rational and irrational magnetic fields},\
    }\href {https://doi.org/10.1103/PhysRevB.14.2239} {\bibfield  {journal}
    {\bibinfo  {journal} {Phys. Rev. B}\ }\textbf {\bibinfo {volume} {14}},\
    \bibinfo {pages} {2239} (\bibinfo {year} {1976})}\BibitemShut {NoStop}%
  \bibitem [{Note1()}]{Note1}%
    \BibitemOpen
    \bibinfo {note} {The strength $\Phi /\phi _0$ is the magnetic flux through a
    unit cell in units of the flux quantum.}\BibitemShut {Stop}%
  \bibitem [{SM()}]{SM}%
    \BibitemOpen
    \href@noop {} {\bibinfo {title} {Supplemental material}}\BibitemShut
    {NoStop}%
  \bibitem [{Note2()}]{Note2}%
    \BibitemOpen
    \bibinfo {note} {$h_+$ decouples with $h_-$ when $t_c=0$, which is helpful
    for theoretical analysis. However, a small coupling $\delta $ will not change
    the results qualitatively.}\BibitemShut {Stop}%
  \bibitem [{\citenamefont {Liu}\ \emph {et~al.}(2016{\natexlab{b}})\citenamefont
    {Liu}, \citenamefont {Zhang},\ and\ \citenamefont
    {Qi}}]{annualreview_liuchaoxing2016}%
    \BibitemOpen
    \bibfield  {author} {\bibinfo {author} {\bibfnamefont {C.-X.}\ \bibnamefont
    {Liu}}, \bibinfo {author} {\bibfnamefont {S.-C.}\ \bibnamefont {Zhang}},\
    and\ \bibinfo {author} {\bibfnamefont {X.-L.}\ \bibnamefont {Qi}},\
    }\bibfield  {title} {\bibinfo {title} {The quantum anomalous hall effect:
    Theory and experiment},\ }\href
    {https://doi.org/10.1146/annurev-conmatphys-031115-011417} {\bibfield
    {journal} {\bibinfo  {journal} {Annu. Rev. Condens. Matter Phys.}\ }\textbf
    {\bibinfo {volume} {7}},\ \bibinfo {pages} {301} (\bibinfo {year}
    {2016}{\natexlab{b}})}\BibitemShut {NoStop}%
  \bibitem [{\citenamefont {Weiße}\ \emph {et~al.}(2006)\citenamefont {Weiße},
    \citenamefont {Wellein}, \citenamefont {Alvermann},\ and\ \citenamefont
    {Fehske}}]{rmp_kpm2006}%
    \BibitemOpen
    \bibfield  {author} {\bibinfo {author} {\bibfnamefont {A.}~\bibnamefont
    {Weiße}}, \bibinfo {author} {\bibfnamefont {G.}~\bibnamefont {Wellein}},
    \bibinfo {author} {\bibfnamefont {A.}~\bibnamefont {Alvermann}},\ and\
    \bibinfo {author} {\bibfnamefont {H.}~\bibnamefont {Fehske}},\ }\bibfield
    {title} {\bibinfo {title} {The kernel polynomial method},\ }\href
    {https://doi.org/10.1103/RevModPhys.78.275} {\bibfield  {journal} {\bibinfo
    {journal} {Rev. Mod. Phys.}\ }\textbf {\bibinfo {volume} {78}},\ \bibinfo
    {pages} {275} (\bibinfo {year} {2006})}\BibitemShut {NoStop}%
  \bibitem [{\citenamefont {MacKinnon}\ and\ \citenamefont
    {Kramer}(1981)}]{prl_mackinnon1981}%
    \BibitemOpen
    \bibfield  {author} {\bibinfo {author} {\bibfnamefont {A.}~\bibnamefont
    {MacKinnon}}\ and\ \bibinfo {author} {\bibfnamefont {B.}~\bibnamefont
    {Kramer}},\ }\bibfield  {title} {\bibinfo {title} {One-parameter scaling of
    localization length and conductance in disordered systems},\ }\href
    {https://doi.org/10.1103/PhysRevLett.47.1546} {\bibfield  {journal} {\bibinfo
     {journal} {Phys. Rev. Lett.}\ }\textbf {\bibinfo {volume} {47}},\ \bibinfo
    {pages} {1546} (\bibinfo {year} {1981})}\BibitemShut {NoStop}%
  \bibitem [{\citenamefont {MacKinnon}\ and\ \citenamefont
    {Kramer}(1983)}]{zpb_mackinnon1983}%
    \BibitemOpen
    \bibfield  {author} {\bibinfo {author} {\bibfnamefont {A.}~\bibnamefont
    {MacKinnon}}\ and\ \bibinfo {author} {\bibfnamefont {B.}~\bibnamefont
    {Kramer}},\ }\bibfield  {title} {\bibinfo {title} {The scaling theory of
    electrons in disordered solids: Additional numerical results},\ }\href
    {https://doi.org/10.1007/BF01578242} {\bibfield  {journal} {\bibinfo
    {journal} {Z. Phys. B}\ }\textbf {\bibinfo {volume} {53}},\ \bibinfo {pages}
    {1} (\bibinfo {year} {1983})}\BibitemShut {NoStop}%
  \bibitem [{\citenamefont {Kramer}\ and\ \citenamefont
    {MacKinnon}(1993)}]{repprogphys_kramer1993}%
    \BibitemOpen
    \bibfield  {author} {\bibinfo {author} {\bibfnamefont {B.}~\bibnamefont
    {Kramer}}\ and\ \bibinfo {author} {\bibfnamefont {A.}~\bibnamefont
    {MacKinnon}},\ }\bibfield  {title} {\bibinfo {title} {Localization: theory
    and experiment},\ }\href {https://doi.org/10.1088/0034-4885/56/12/001}
    {\bibfield  {journal} {\bibinfo  {journal} {Rep. Prog. Phys.}\ }\textbf
    {\bibinfo {volume} {56}},\ \bibinfo {pages} {1469} (\bibinfo {year}
    {1993})}\BibitemShut {NoStop}%
  \bibitem [{\citenamefont {Slevin}\ and\ \citenamefont
    {Ohtsuki}(2014)}]{njp_tomi2014}%
    \BibitemOpen
    \bibfield  {author} {\bibinfo {author} {\bibfnamefont {K.}~\bibnamefont
    {Slevin}}\ and\ \bibinfo {author} {\bibfnamefont {T.}~\bibnamefont
    {Ohtsuki}},\ }\bibfield  {title} {\bibinfo {title} {Critical exponent for the
    anderson transition in the three-dimensional orthogonal universality class},\
    }\href {https://doi.org/10.1088/1367-2630/16/1/015012} {\bibfield  {journal}
    {\bibinfo  {journal} {New J. Phys.}\ }\textbf {\bibinfo {volume} {16}},\
    \bibinfo {pages} {015012} (\bibinfo {year} {2014})}\BibitemShut {NoStop}%
  \bibitem [{\citenamefont {Zhang}\ \emph {et~al.}(2021)\citenamefont {Zhang},
    \citenamefont {Chen}, \citenamefont {Wu}, \citenamefont {Jiang},
    \citenamefont {Liu}, \citenamefont {Sun},\ and\ \citenamefont
    {Xie}}]{prb_zhangzhiqiang2021}%
    \BibitemOpen
    \bibfield  {author} {\bibinfo {author} {\bibfnamefont {Z.-Q.}\ \bibnamefont
    {Zhang}}, \bibinfo {author} {\bibfnamefont {C.-Z.}\ \bibnamefont {Chen}},
    \bibinfo {author} {\bibfnamefont {Y.}~\bibnamefont {Wu}}, \bibinfo {author}
    {\bibfnamefont {H.}~\bibnamefont {Jiang}}, \bibinfo {author} {\bibfnamefont
    {J.}~\bibnamefont {Liu}}, \bibinfo {author} {\bibfnamefont {Q.-f.}\
    \bibnamefont {Sun}},\ and\ \bibinfo {author} {\bibfnamefont {X.~C.}\
    \bibnamefont {Xie}},\ }\bibfield  {title} {\bibinfo {title} {Chiral interface
    states and related quantized transport in disordered chern insulators},\
    }\href {https://doi.org/10.1103/PhysRevB.103.075434} {\bibfield  {journal}
    {\bibinfo  {journal} {Phys. Rev. B}\ }\textbf {\bibinfo {volume} {103}},\
    \bibinfo {pages} {075434} (\bibinfo {year} {2021})}\BibitemShut {NoStop}%
  \end{thebibliography}
\end{document}